\documentclass{moriond}

\usepackage{float}

\def\Journal#1#2#3#4{{#1} {\bf #2}, #3 (#4)}

\def\NIMA{{\em Nucl. Instrum. Methods} A}

\def\PRL{\em Phys. Rev. Lett.}
\def\PRD{{\em Phys. Rev.} D}

\def\APJ{{\em ApJ.}}

\def\be{\begin{equation}}
\def\ee{\end{equation}}
\def\bea{\begin{eqnarray}}
\def\eea{\end{eqnarray}}

\newcommand{\Photo}{\includegraphics[height=35mm]{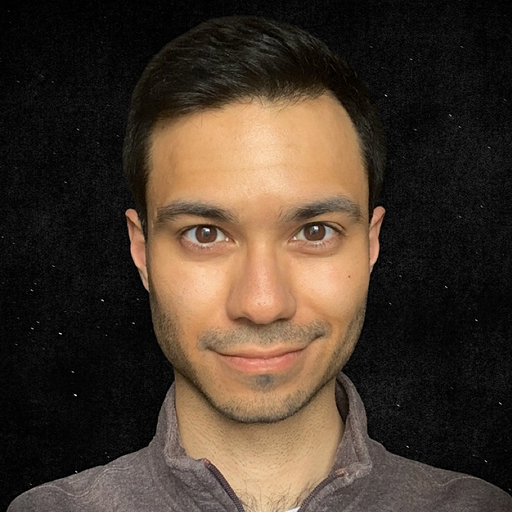}}

\begin{document}
\vspace*{4cm}
\title{Latest results from Super-Kamiokande}

\author{ A.D. Santos \\ (on behalf of the Super-Kamiokande collaboration)}

\address{\'Ecole Polytechnique, IN2P3-CNRS, Laboratoire Leprince-Ringuet, \\ F-91120 Palaiseau, France}

\maketitle\abstracts{Super-Kamiokande is the world's largest water Cherenkov experiment with its 50-kton tank of ultrapure water, recently doped with gadolinium to enhance neutron capture identification. It is a highly versatile, multi-purpose experiment in the MeV-TeV range, and here we will summarize the latest results and advancements in the atmospheric-$\nu$, solar-$\nu$, and Diffuse Supernova Neutrino Background analyses.}

\section{Introduction}

The Super-Kamiokande (Super-K/SK) experiment is a 50-kton water Cherenkov detector located in the Kamioka mine in Japan. It is shielded by an overburden of around 1 km of rock (2.7 km water equivalent) to reduce contamination from cosmic ray muons created in the Earth's atmosphere. It comprises a 32-kton inner detector lined with more than 11,000 20-inch photomultiplier tubes (PMTs) and an 18-kton outer detector primarily serving as a muon veto with nearly 2,000 8-inch PMTs. It has been in operation since April 1996. Super-K phases I to III reached until August 2008, and the SK-IV period from September 2008 to May 2018 saw a front-end electronics upgrade allowing for a larger timing window for saving PMT data and, therefore, for neutron capture identification. Following SK-IV, the detector was prepared for the addition of gadolinium into the tank to enhance this neutron ``tagging." At present, Super-K has undergone two gadolinium loading phases in 2020 and 2022, resulting in 0.01\% and 0.03\% Gd mass concentration, respectively.\cite{gd22,gd24} These levels already correspond to 50\% and 75\% of neutron captures occurring on Gd. Recent physics results of Super-K cover a wide variety of topics, including atmospheric neutrinos,\cite{atm24,ncqe24} solar neutrinos,\cite{solvar23,solres23} supernova neutrinos,\cite{sn22,snsearch22,sn24} the Diffuse Supernova Neutrino Background (DSNB),\cite{dsnb21,dsnb23} proton decay,\cite{pd22} and dark matter searches.\cite{dm23} Here, we highlight a few of these latest results and advancements for the atmospheric, solar, and DSNB analyses.

\section{Atmospheric Neutrinos}

Cosmic ray interactions with nuclei in the Earth's atmosphere generate hadronic showers whose final states include neutrinos and anti-neutrinos of electron and muon flavor. These neutrinos span orders of magnitude in energy, starting at $\mathcal{O}$(100 MeV). Super-K can detect them from any direction, probing flavor oscillations covering baselines of about 15 km to 13000 km. Moreover, atmospheric neutrinos passing through the Mikheyev-Smirnov-Wolfenstein (MSW) resonance in the Earth's interior add another dimension of flavor conversion. The oscillation effects of a CP-violating Dirac phase are dominant at sub-GeV energies, while those of different mass orderings are concentrated at the multi-GeV scale. With this all in mind, discriminating between $e$-/$\mu$-like events with a $\nu$/$\bar{\nu}$ separation is crucial for extracting physics from atmospheric neutrinos.

\begin{figure}[H]
\begin{minipage}{0.35\linewidth}
\centerline{\includegraphics[width=\linewidth]{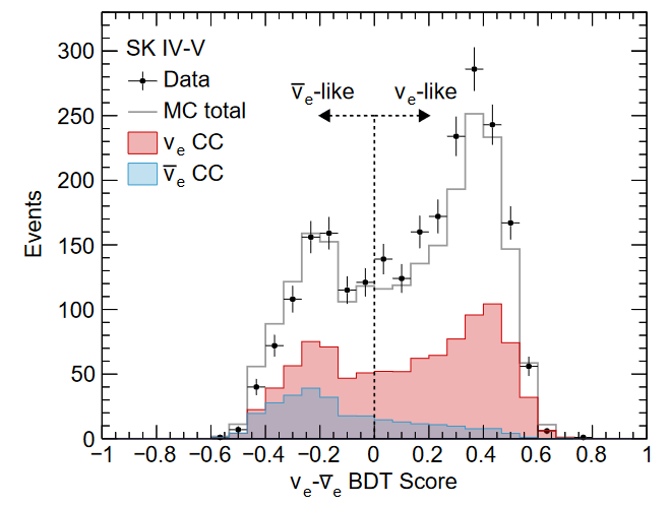}}
\end{minipage}
\hfill
\begin{minipage}{0.6\linewidth}
\centerline{\includegraphics[width=\linewidth]{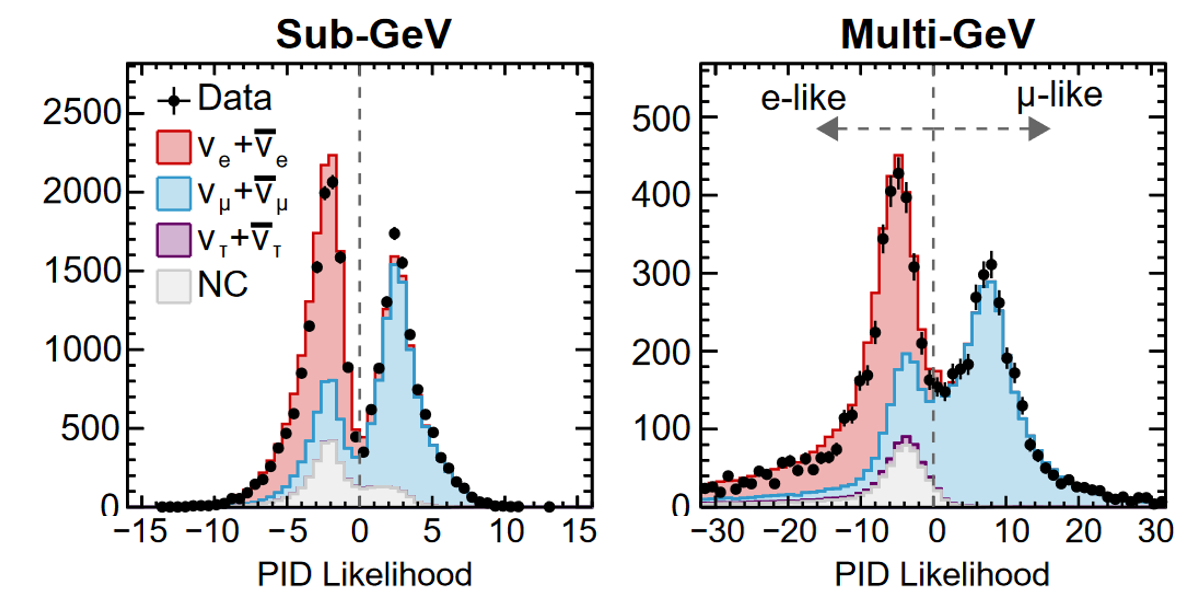}}
\end{minipage}
\caption[]{(left) Atmospheric $\nu_e$-$\bar{\nu}_e$ separation using a BDT discriminator and (right) atmospheric $\nu$ flavor separation for sub-GeV and multi-GeV samples.}
\label{fig:atm_perf}
\end{figure}

Super-K recently published results from data-taking phases I-V.\cite{atm24} An expanded fiducial volume was adopted, integrating events with reconstructed distances down to as little as 100 cm away from the detector walls. This represents a 20\% increase in events compared to the previous 200 cm cutoff. Another notable improvement is the use of neutron tagging. Since anti-neutrino interactions such as $\bar{\nu}_l + p \to l^+ + n + X$ produce neutrons, searching for neutron captures in the water enhances $\nu/\bar{\nu}$ separation. Further separation is done with a Boosted Decision Tree, and flavor identification is possible by exploiting the geometry of the Cherenkov rings. The performance of these approaches in shown in Figure \ref{fig:atm_perf}.

\begin{figure}[H]
\begin{minipage}{0.34\linewidth}
\centerline{\includegraphics[width=\linewidth]{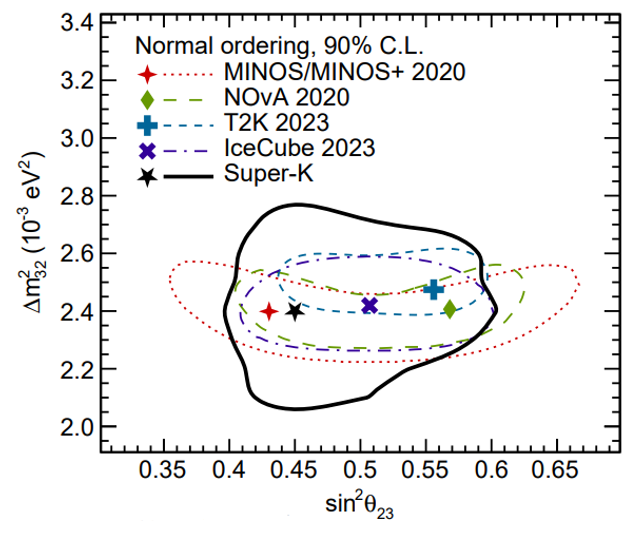}}
\end{minipage}
\hfill
\begin{minipage}{0.65\linewidth}
\centerline{\includegraphics[width=\linewidth]{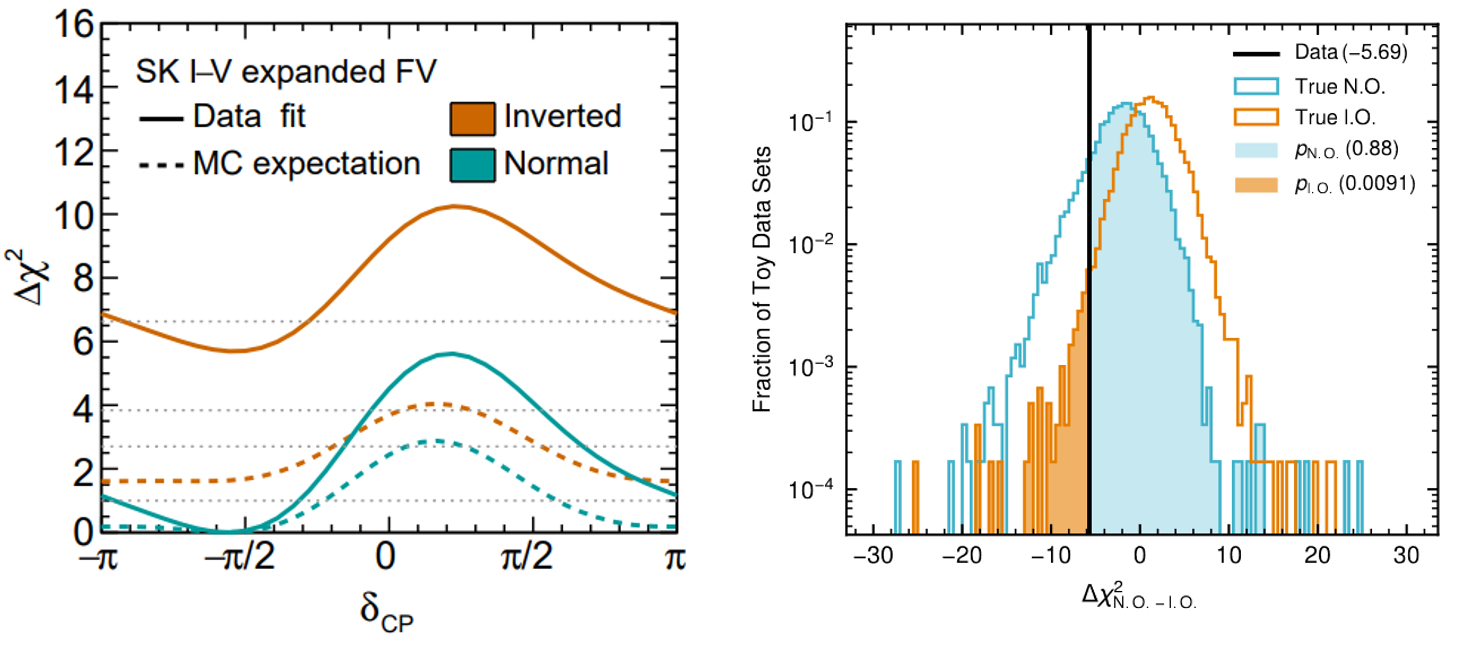}}
\end{minipage}
\caption[]{(left) Best-fit contours for $\Delta m_{32}^2$ and $\sin^2(\theta_{23})$ for SK (black, solid) compared to other experiments, (middle) $\Delta \chi^2$ for various $\delta_{CP}$ values, and (right) distributions in $\Delta \chi_{(N.O-I.O)}^2$ under the true N.O./I.O. hypotheses with the observed data (black, vertical).}
\label{fig:atm_res}
\end{figure}

The numerical results are summarized in Table \ref{tab:atm_res}. The best-fit $\Delta m_{32}^2$ value is consistent with other experimental measurements, and the best-fit $\sin^2(\theta_{23})$ value lies in the first octant (i.e., $\sin^2(\theta_{23}) < 0.5$). CP-conservation is lightly disfavored with a best-fit value around $-\pi/2$. This is a maximally CP-violating phase and is consistent with the $\delta_{CP}$ measurement coming from T2K.\cite{t2k20} Finally, we examine the mass ordering determination using the CL$_s$ method\cite{cls02} in which CL$_s \equiv p_{I.O}/(1 - p_{N.O.})$ here. The two p-values are calculated from the $\Delta \chi_{(N.O-I.O)}^2$ distributions under each mass ordering hypothesis. With $p_{I.O} = 0.0091$ and $p_{N.O.} = 0.88$, this gives CL$_s = 0.077$, a rejection of inverted ordering at the 92.3\% confidence level. All of these results are also shown in Figure \ref{fig:atm_res}.

\begin{table}[H]
\caption[]{Best-fit values for the oscillation parameters with $\pm1\sigma$ allowed regions assuming a $\chi^2$ distribution with one degree of freedom. The reactor constraint $\sin^2 \theta_{13} = 0.0220\pm0.0007$ is imposed.}
\label{tab:atm_res}
\vspace{0.4cm}
\begin{center}
\begin{tabular}{|c|c|c|c|c|}
\hline
& & & & \\
Mass Ordering & $\left| \Delta m_{32}^2 \right|$ ($10^{-3}$ eV$^2$) & $\sin^2(\theta_{23})$ & $\delta_{CP}$ ($-\pi$, $\pi$) & $\Delta \chi_{(N.O-I.O)}^2$ \\
\hline
Normal & $2.40_{-0.09}^{+0.07}$ & $0.45_{-0.03}^{+0.06}$ & $-1.75_{-1.25}^{+0.76}$ &  \\ 
& & & & $-5.69$ \\ 
Inverted & $2.40_{-0.12}^{+0.06}$ & $0.45_{-0.03}^{+0.08}$ & $-1.75_{-1.22}^{+0.89}$ &  \\ 
\hline
\end{tabular}
\end{center}
\end{table}

\section{Solar Neutrinos}

\begin{figure}[H]
\begin{minipage}{0.49\linewidth}
\centerline{\includegraphics[width=\linewidth]{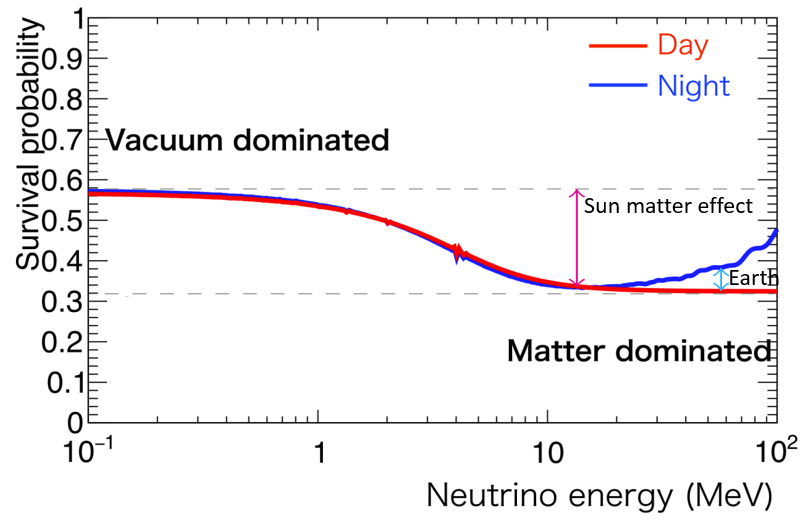}}
\end{minipage}
\hfill
\begin{minipage}{0.47\linewidth}
\centerline{\includegraphics[width=\linewidth]{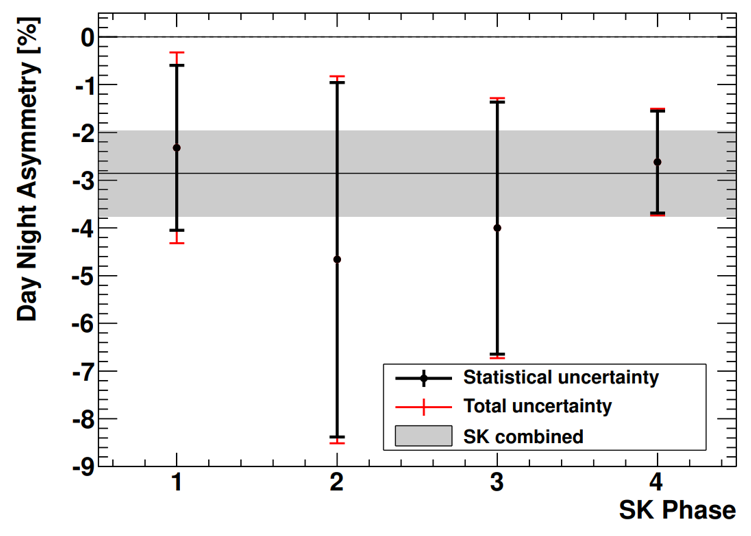}}
\end{minipage}
\caption[]{(left) Solar $\nu_e$ survival probability as a function of energy, day/night periods and (right) SK observed day/night solar $\nu_e$ asymmetry for each phase (I, II, III, IV) and globally (gray band).}
\label{fig:solar_daynight}
\end{figure}

Solar neutrinos are produced from several different nuclear fusion channels in the Sun. While Super-K is sensitive to $^8$B neutrinos, other processes include $pp$, $pep$, $^7$Be, $hep$, and CNO cycle neutrinos. The centerpiece of solar neutrino oscillation physics is the MSW effect. After $\nu_e$ are produced in dense solar matter, they propagate as pure matter eigenstates, different from the usual vacuum eigenstates. Assuming adiabatic crossing through the solar MSW resonance, those above $\mathcal{O}$(1 MeV) arrive at the vacuum of space as purely $\nu_2$ mass eigenstates. They are then detected through electron elastic scattering in Super-K for which the $\nu_e$ channels dominate. For those passing through the Earth before arriving at the detector, an additional regeneration of electron flavor is predicted to occur from another MSW resonance. These two effects are illustrated in Figure \ref{fig:solar_daynight}.

Super-K has now released its latest solar neutrino results through the SK-IV period.\cite{solres23} Several improvements to the analysis were implemented. First, the data acquisition (PMT hit) threshold was lowered in May 2015, notably increasing the efficiency of events in [3.49, 3.99] MeV in the MSW transition region. Next, cosmic ray muon spallation backgrounds were better reduced. While several reduction steps are already in use (and others are improved), the addition of searching for clusters of neutrons from these backgrounds removes 54\% of spallation with only a 1.4\% loss in signal efficiency. Finally, the time evolution of the PMT gain was used for better energy reconstruction. In all, the SK-IV solar neutrino count in [3.49, 19.49] MeV was $65,443_{-388}^{+388}$ (stat.)$\pm 925$ (sys.).

\begin{figure}[H]
\begin{minipage}{0.48\linewidth}
\centerline{\includegraphics[width=\linewidth]{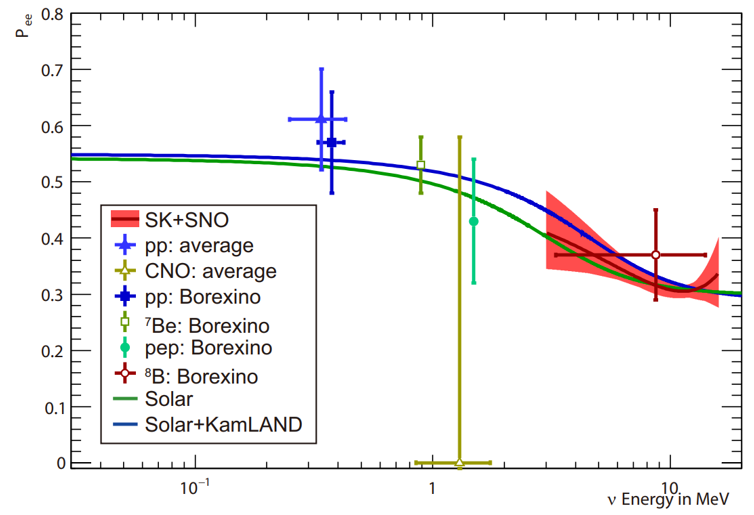}}
\end{minipage}
\hfill
\begin{minipage}{0.47\linewidth}
\centerline{\includegraphics[width=\linewidth]{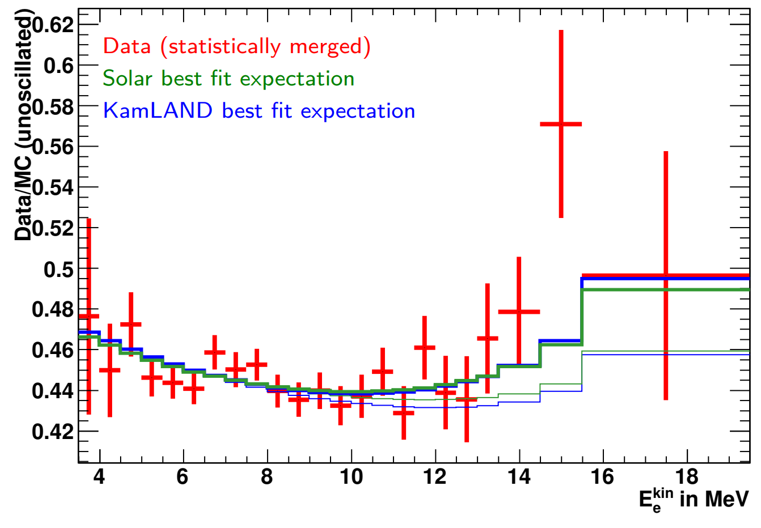}}
\end{minipage}
\caption[]{(left) SK+SNO fitted solar $\nu_e$ survival probability as a function of energy and (right) binned data for the fitted solar $\nu_e$ survival probability.}
\label{fig:solar_upturn}
\end{figure}

\begin{figure}[H]
\centering
\begin{minipage}{0.8\linewidth}
\centerline{\includegraphics[width=\linewidth]{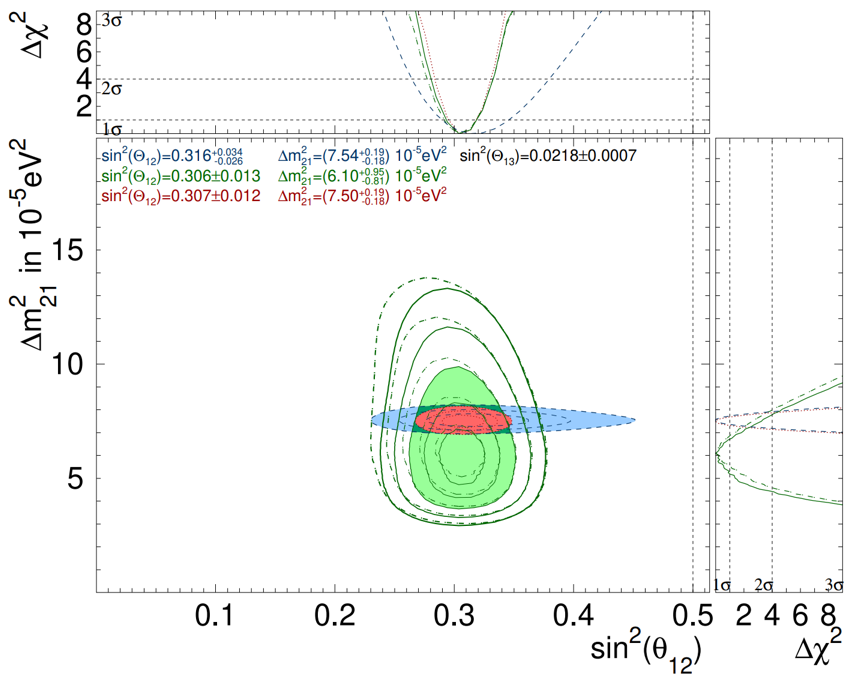}}
\end{minipage}
\caption[]{Best-fit contours for $\Delta m_{21}^2$ and $\sin^2(\theta_{12})$ for the global solar fit (green, solid), the KamLAND experiment (blue), and the solar+KamLAND combination (red).}
\label{fig:solar_2D_res}
\end{figure}

The latest results demonstrate the first observation of a day/night asymmetry at the $3\sigma$ level when combining SK phases I-IV (see Figure \ref{fig:solar_daynight}). Probing the so-called solar ``upturn" in $\nu_e$ survival probability while descending in neutrino energy (Figure \ref{fig:solar_upturn}), the data were fit to exponential, quadratic, and cubic functions. A distorted spectrum was favored at 2.1$\sigma$ for SK+SNO data, which hints at an observation of the upturn. Finally, Figure \ref{fig:solar_2D_res} shows the best-fit contours for $\Delta m_{21}^2$ and $\sin^2(\theta_{12})$. The global solar neutrino fit is more sensitive to $\sin^2(\theta_{12})$ than the KamLAND reactor anti-neutrino analysis, while KamLAND is more sensitive to the mass-squared splitting. Assuming CPT invariance (and the 3-flavor oscillation framework), both solar $\nu_e$ and reactor $\bar{\nu}_e$ experiments should observe the same mixing parameters. A 1.5$\sigma$ tension is observed in the best-fit mass-squared splitting values at this time.

\section{Supernova Neutrinos and the DSNB}

Core-collapse supernovae (CCSNe) are violent explosions at the end of massive stars' lives, resulting in either a neutron star or black hole. During the explosion, 99\% of the energy is released in the form of neutrinos. Since neutrinos can more easily escape the optically opaque environment than photons, there is a chance to have advance notice in neutrinos before optical observation. Within the Milky Way Galaxy, CCSNe take place maybe once per century, and Super-K is prepared with a pre-SN alarm in the event of a galactic SN.\cite{sn22,sn24}  Throughout the entire observable universe, though, a CCSN occurs about once every second. This should result in a constant, isotropic flux of redshifted SN neutrinos known as the Diffuse Supernova Neutrino Background (DSNB). 

While it has yet to be observed, Super-K has the world's leading sensitivity to the DSNB through the inverse beta decay (IBD) channel. The experimental signature is then the coincidence of a prompt positron followed by a neutron capture. At energies below 16 MeV, spallation backgrounds and reactor neutrinos begin to overwhelm the DSNB flux. Beyond 30~MeV, atmospheric neutrino charged-current interactions dominate. Atmospheric neutral current quasi-elastic (NCQE) interactions are problematic below 20 MeV.

\begin{figure}[H]
\begin{minipage}{0.46\linewidth}
\centerline{\includegraphics[width=\linewidth]{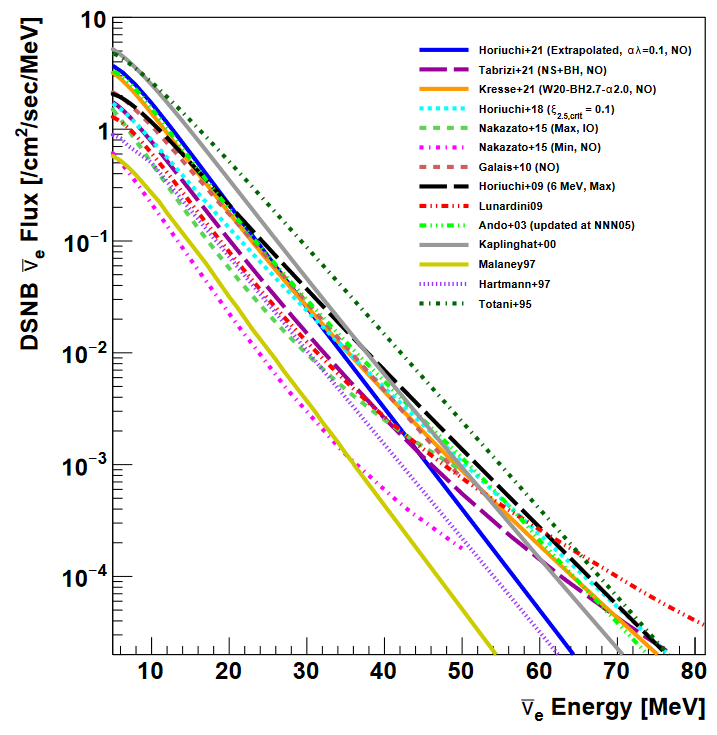}}
\end{minipage}
\hfill
\begin{minipage}{0.5\linewidth}
\centerline{\includegraphics[width=\linewidth]{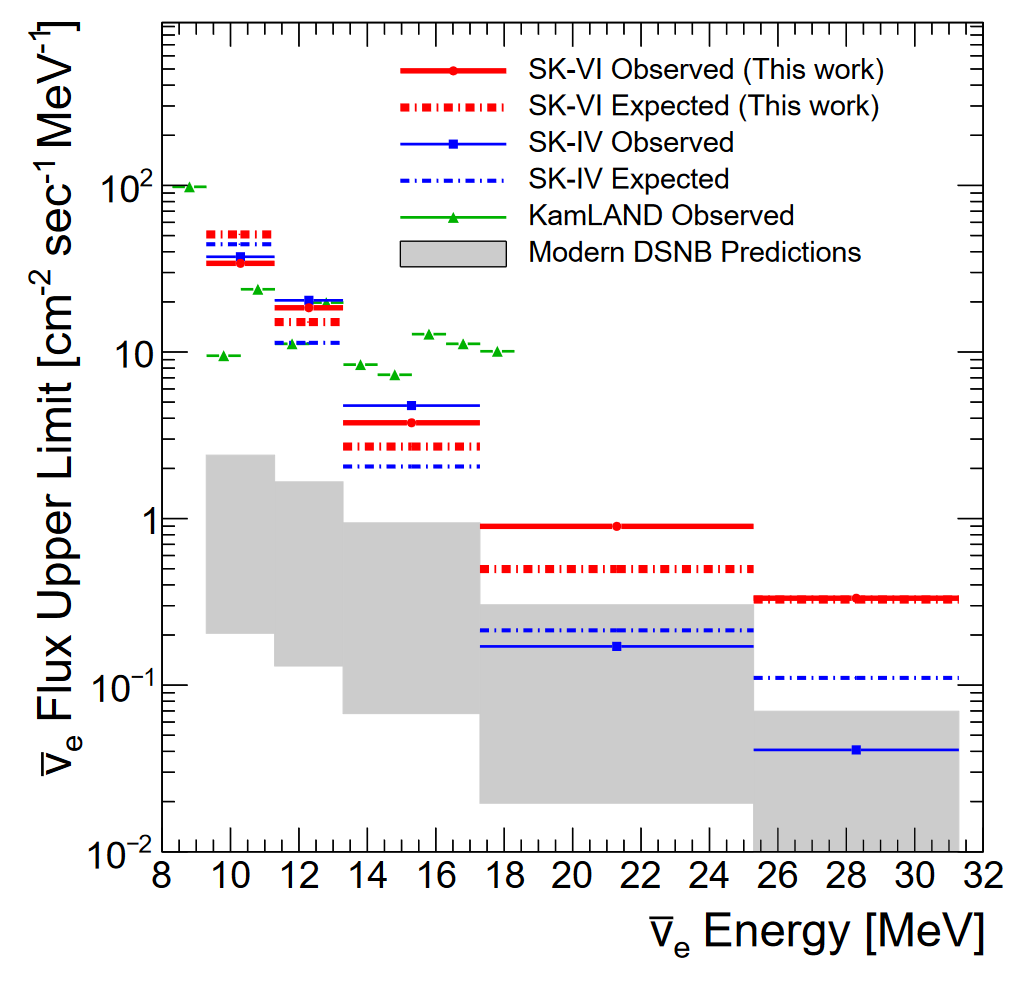}}
\end{minipage}
\caption[]{(left) Selected DSNB theoretical models and (right) 90\% C.L. upper limits with expected sensitivity for SK-IV, SK-VI.}
\label{fig:dsnb_limits}
\end{figure}

\begin{figure}[H]
\begin{minipage}{\linewidth}
\centerline{\includegraphics[width=\linewidth]{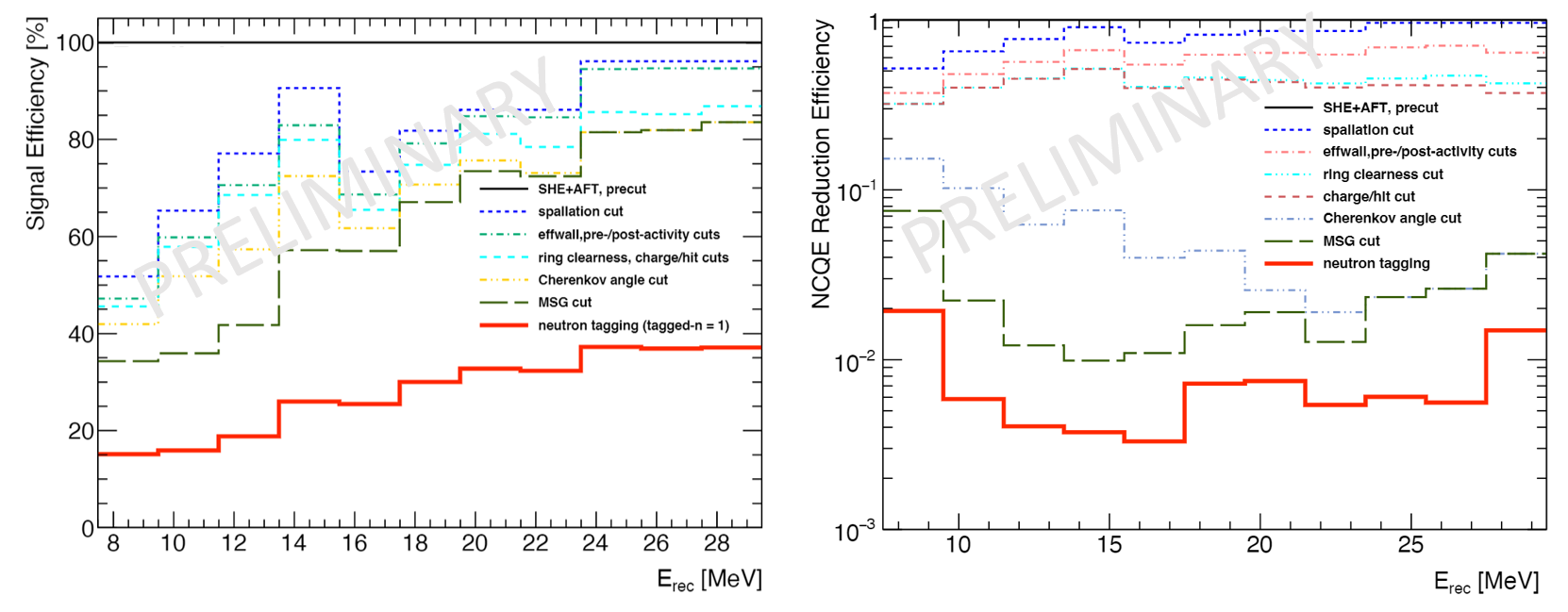}}
\end{minipage}
\caption[]{(left) Inverse beta decay signal efficiency and (right) atmospheric neutrino NCQE background acceptance after successive reduction steps in the SK-VI DSNB analysis.}
\label{fig:dsnb_eff}
\end{figure}

The tightest upper limits on the DSNB are especially driven by the 2970.1 livedays of data taken from the SK-IV period.\cite{dsnb21} A more recent study included 552.2 livedays from the SK-VI period, the first with Gd-loading (0.01\%).\cite{dsnb23} The observed upper limits compared to a wide range of DSNB theoretical models are shown in Figure \ref{fig:dsnb_limits}. A more in-depth, multi-phase DSNB analysis is underway to fully include SK-Gd data. There are several improvements. First, the use of SK-Gd data significantly enhances the neutron capture signal. To exploit this, dedicated machine learning algorithms for neutron tagging are applied to the event selection. Next, as with the SK I-IV DSNB analysis, a spectral fit is performed to simultaneously fit signal and background regions. Finally, a new reduction for atmospheric NCQE interactions is introduced. It makes use of the ``multiple scattering goodness" (MSG) variable already employed in the SK solar neutrino analysis and brings the multi-cone NCQE backgrounds down to the $\mathcal{O}$(1\%) level. The energy-dependent IBD signal efficiency and NCQE background acceptance rates are shown in Figure \ref{fig:dsnb_eff}.

\section{Conclusion}

The Super-Kamiokande experiment continues to produce results for a vast array of physics topics. The latest atmospheric neutrino analysis suggests the presence of a CP-violating Dirac phase along with a 92.3\% confidence level rejection of the inverted mass ordering. The solar neutrino results shown here provide the first evidence at 3$\sigma$ for a day/night asymmetry in solar $\nu_e$ flux driven by an MSW resonance in the Earth, and the SK+SNO combined analysis favors the observation of the solar ``upturn" in $\nu_e$ survival probability at 2.1$\sigma$. Finally, the Super-K DSNB analysis produces the world's tightest limits with new results already underway for the SK-Gd era using more sophisticated neutron tagging, spectral fitting, and atmospheric background reduction. 

\section*{Acknowledgments}

We gratefully acknowledge cooperation of the Kamioka Mining and Smelting Company.
The Super-Kamiokande experiment was built and has been operated with funding from the
Japanese Ministry of Education, Science, Sports and Culture, 
and the U.S. Department of Energy.

\end{document}